\documentclass[aip,amsmath,amssymb,reprint]{revtex4-1}
\usepackage{mathptmx}
\usepackage[T1]{fontenc}
\usepackage[utf8]{inputenc}
\usepackage{amsmath}
\usepackage{graphicx}

\makeatletter

\providecommand{\tabularnewline}{\\}

\usepackage{dcolumn}% Align table columns on decimal point
\usepackage{bm}% bold math

\makeatother

\begin{document}

\preprint{AIP/123-QED}

\title{Thermal Collapse of a Skyrmion}

\author{Amel Derras-Chouk}

\author{Eugene M. Chudnovsky}

\author{Dmitry A. Garanin}

\affiliation{Physics Department, Herbert H. Lehman College and Graduate School,
The City University of New York ~\\
 250 Bedford Park Boulevard West, Bronx, New York 10468-1589, USA}

\date{\today}
\begin{abstract}
Thermal collapse of an isolated skyrmion on a two-dimensional spin
lattice has been investigated. The method is based upon solution of
the system of stochastic Landau-Lifshitz equations for up to $10^{4}$
spins. The recently developed pulse-noise algorithm has been used
for the stochastic component of the equations. The collapse rate follows
the Arrhenius law. Analytical formulas derived within a continuous
spin-field model support numerically-obtained values of the energy
barrier. The pre-exponential factor is independent of the phenomenological
damping constant that implies that the skyrmion is overcoming the
energy barrier due to the energy exchange with the rest of the spin
system. Our findings agree with experiments, as well as with recent
numerical results obtained by other methods. 
\end{abstract}
\maketitle

\section{Introduction}

Skyrmions are whirls of spins stabilized by topology. The topological
protection of skyrmions has motivated recent research on their nucleation
and manipulation in thin magnetic films. It arises from the mapping
of a three-component fixed-length spin field on a two-dimensional
(2D) coordinate space \cite{ec-book,Brown-book,Manton-book}. The
topologically protected metastable skyrmions solutions for the 2D
Heisenberg exchange model in the continuous approximation obtained
by Belavin and Polyakov (BP) \cite{BelPolJETP75} are scale invariant,
that is, their energy is independent of their size. In practice, scale
invariance is violated by the discreteness of the atomic lattice \cite{cai12},
as well as by other interactions such as Zeeman interaction, dipole-dipole
interaction (DDI), perpendicular magnetic anisotropy (PMA), interaction
of skyrmions with defects and boundaries, etc. This leads to the stabilization
of the skyrmion size at some value minimizing the total energy, or,
to the contrary, to their collapse or expansion followed by the transformation
into the domain structure. 

Currently, there is a great interest in skyrmions in non-centrosymmetric
materials. The lack of inversion symmetry results in the Dzyaloshinskii-Moriya
interaction (DMI). It competes with other material-dependent interactions,
providing metastability or even thermodynamic stability of skyrmions
within a certain area of the phase diagram \cite{Leonov-NJP2016,buttner18}.
Stable isolated skyrmions have been experimentally observed \cite{boulle16,moreau16,Fert-Nature2017}
even at room temperatures.

Skyrmions provide a promising avenue for new forms of memory storage
because information can be encoded in them as bits \cite{Nagaosa2013,Tomasello,Zhang2015,Klaui2016,Leonov-NJP2016,Hoffmann-PhysRep2017,Fert-Nature2017}.
Recent advances in nucleation methods have accelerated interest in
skyrmions by demonstrating the feasibility of skyrmion writing devices.
One promising method involves application of a spin-polarized current
using a scanning tunneling microscope. Such a method can both nucleate
and erase skyrmions, as was demonstrated on a PbFe/Ir(111) system
\cite{romming13}. Additionally, recent experimental and theoretical
work has explored the possibility of skyrmion creation by temperature
\cite{Zhang2018}, by local heating \cite{koshibae14,Berruto-PRL2018}
and with the help of the magnetic force microscope (MFM) tip \cite{Senfu-APL2018}
or of a magnetic dipole \cite{garanin-dipole}. The existence of skyrmions
depends on both the material parameters and the external conditions,
such as the magnetic field and temperature. It has been experimentally
demonstrated that the size of a skyrmion can be tuned by the external
field, with its radius shrinking on increasing the field opposite
to the direction of spins in the skyrmion until the skyrmion disappears
\cite{romming13,romming15}.

The longevity of metastable skyrmions with respect to thermal or quantum
collapse (also called thermal or quantum stability) is important for
potential applications. Quantum decay of a skyrmion has been recently
studied by a method based upon finding an instanton solution of the
imaginary-time equations of motion \cite{quantum}. The statistical
mechanics problem of the thermal collapse of a skyrmion is more involved.
It has been explored for various systems, both for a skyrmion in a
racetrack \cite{cortes17,bessarab18} and an isolated skyrmion in
a thin film \cite{stosic17,rohart16,desplat18,hagemeister15,rozsa16}.
Although most of the investigators have studied purely 2D systems
\cite{cortes17,rohart16,hagemeister15,varentsova18,desplat18,siemens16,bessarab18},
some have considered multilayered systems as well \cite{stosic17}.
The latter are relevant to recent experimental advances in creating
room temperature skyrmions, while numerical models of purely 2D systems
apply to skyrmions found in B20 materials \cite{koshibae17}. Some
previous work uses Monte Carlo simulations \cite{hagemeister15,siemens16,rozsa16}
that capture the main physics of the thermally-activated decay but
cannot establish its overall time scale for the lack of connection
to the actual dynamics. There are semi-analytical methods that search
for the minimum-energy path out of the metastable state to find the
energy barrier \cite{bessarab18,stosic17,desplat18,vonmalottki17,varentsova18,cortes17,rohart16}.
One of these methods is the Nudged Elastic Band Method (NEBM) \cite{bessarab18,stosic17,desplat18,vonmalottki17,varentsova18,cortes17},
that is efficient in the cases when the decay path is not obvious
since it is mediated by the proximity of the skyrmion to the defects
or to the boundaries. After the decay path is found, the skyrmion
decay rate is usually written in the form of the Arrhenius expression
with the exponent defined by the energy barrier and the prefactor
given by the Langer's formula \cite{desplat18}.

In this paper, we present the results of the lifetime computations
for an isolated metastable skyrmion using the Landau-Lifshitz equation
\cite{LandauLifshitz35} for a system of classical spins augmented
by the stochastic Langevin fields, the so-called Landau-Lifshitz-Langevin
(LLL) equation. In comparison to the Monte Carlo simulations, this
method is more realistic as it describes the actual dynamics of the
system. We consider skyrmions in an pure infinite 2D system, modeled
by a finite system with periodic boundary conditions. In this case,
one can estimate the energy barrier analytically to compare with the
numerical results without a need to use NEMB and similar involved
numerical methods. 

For a single spin, as well as for a single particle, the prefactor
in the Arrhenius escape rate nontrivially depends on the damping constant
that accounts for the coupling with the bath, considered phenomenologically.
There are regimes of high, intermediate, and weak damping, first found
by Kramers \cite{kramers1940} for the particle and later established
for the spin by Brown \cite{brown1963} and later workers (see, e.g.,
Refs. \cite{Coffee-Kalmykov-book,byrne2019}). The stochastic equation
of motion for the particle or spin is equivalent to the corresponding
Fokker-Planck equation that can be solved analytically or numerically.
In particular, at low temperatures the decay rate is controlled by
the lowest eigenvalue of the Fokker-Planck operator.

The situation is different for many-body systems such as spin systems
with skyrmions. First, the multidimensional Fokker-Planck equation
cannot be solved numerically (although the analytical Langer's approach
is still possible under the assumption of a high-to-intermediate damping
\cite{desplat18}). Thus the direct solution of the LLL equation for
a system of spins is the only available numerical method, quite feasible
with the modern computing power (for a recent reference, see Ref.
\cite{evans2018Springer}). Second, the skyrmion can exchange its
energy with the rest of the spin system, in particular, with the reservoire
of spin waves. The thermal energy of the system is proportional to
its size and exceeds by far the energy needed for a skyrmion to overcome
the barrier and collapse. Thus, it is not obvious whether the phenomenological
bath is playing any role in the dynamics in the realistic low-damping
case, apart from providing the temperature to the systems via the
balance of excitation and damping. Indeed, our results show that the
prefactor in the skyrmion collapse rate is independent of the damping
constant $\alpha$, the so-called Gilbert constant.

In this paper, the LLL equation is treated with the help of the recently
developed pulse-noise approach \cite{garanin17,garanin18} that replaces
the continuous white noise by discrete noise pulses between which
the dynamics of the spin system can be computed by a standard high-order
ordinary-differential-equation (ODE) solvers. This approach works
well in the case of low damping typical for spin systems, and it reproduces
the results of the LLL equation solved by standard methods, however,
with the computational speed of the noiseless dynamics. At low temperatures,
the results fit the Arrhenius law, typical for overbarrier transitions
driven by temperature \cite{bessarab18,stosic17,desplat18,vonmalottki17,varentsova18,cortes17,rohart16}.
Indeed, the Arrhenius law has been observed experimentally in the
decay of the array of skyrmions in a film \cite{wild17}.

The paper is organized as follows. In Section \ref{model} the analytical
model of the static properties of the skyrmion based on the BP skyrmion
shape is discussed. In Section \ref{numerical-methods}, the dynamics
is introduced and the numerical method are presented. In Section \ref{thermal-collapse},
results of the numerical computation and comparison to the analytical
model are given. Lastly, a discussion of the results is done in Section
\ref{discussion}.

%%%%%%%%%%%%%%%%%%%%%%%%%%%%%%%%%%%%%%%%%%%%%%%%%%%%%%%%%%%%%%%%%%%%%%%%%% The model

\section{The model}

\label{model}

We consider a two-dimensional square lattice of spins considered as
normalized three-component vectors, $\textbf{s}_{i}\equiv\textbf{S}_{i}/S$,
where $S$ is the atomic spin value and $i$ refers to the lattice
site. The Hamiltonian of the system is given by 
\begin{align}
\mathcal{H}= & -\frac{S^{2}}{2}\sum_{ij}J_{ij}\textbf{s}_{i}\cdot\textbf{s}_{j}\nonumber \\
 & -S^{2}A\sum_{i}(\textbf{s}_{i}\times\textbf{s}_{i+\hat{x}})\cdot\hat{x}+(\textbf{s}_{i}\times\textbf{s}_{i+\hat{y}})\cdot\hat{y}\nonumber \\
 & -S\textbf{H}\cdot\sum_{i}\textbf{s}_{i}.\label{energy-discrete}
\end{align}
The first term represents the Heisenberg exchange energy with the
exchange constant $J$ between the nearest neighbors. The second term
represents the Dzyaloshinskii-Moriya interaction (DMI). For certainty
we have chosen a Bloch type DMI that favors the Bloch-type skyrmions
with the chirality angle $\gamma=\pi/2$ (see below) for $A>0$. The
last term is the Zeeman interaction energy due to the external magnetic
field induction $\mathbf{B}$, and $\mathbf{H}\equiv g\mu_{B}\mathbf{B}$
is the magnetic field in the energy units (with $\mu_{B}$ being the
Bohr magneton and $g$ being the gyromagnetic factor). The skyrmion
that we consider will be stabilized by a field applied perpendicular
to the plane so that the Zeeman term becomes $-SH\sum_{i}s_{i,z}$.

The continuous analog of this energy is 
\begin{align}
{\cal H} & =\frac{1}{2}JS^{2}\int dxdy\left[\left(\frac{\partial{\bf s}}{\partial x}\right)^{2}+\left(\frac{\partial{\bf s}}{\partial y}\right)^{2}\right]\nonumber \\
 & -\frac{1}{24}JS^{2}\int dxdy\left[\left(\frac{\partial^{2}{\bf s}}{\partial x^{2}}\right)^{2}+\left(\frac{\partial^{2}{\bf s}}{\partial y^{2}}\right)^{2}\right]\nonumber \\
 & +AS^{2}\int dxdy\left[\left({\bf s}\times\frac{\partial{\bf s}}{\partial x}\right)\cdot\hat{x}+\left({\bf s}\times\frac{\partial{\bf s}}{\partial y}\right)\cdot\hat{y}\right]\nonumber \\
 & -HS\int dxdy\,s_{z},\label{E-continuous}
\end{align}
where all lengths are measured in the units of the lattice spacing
$a$. The second term in this expression arises from taking into account
the next derivatives in the expansion of the discrete form of the
exchange energy that becomes important for small-size skyrmions creating
large gradients of the spin field.

The continuous unit-length spin field ${\bf s}$ is characterized
by the topological charge: 
\begin{equation}
Q=\frac{1}{4\pi}\int dxdy\hspace{5pt}\textbf{s}\cdot\bigg(\frac{\partial\textbf{s}}{\partial x}\times\frac{\partial\textbf{s}}{\partial y}\bigg).\label{Q}
\end{equation}
that takes discrete values $Q=0,\pm1,\pm2,...$

The first, dominant, term in Eq.\ (\ref{E-continuous}) gives rise
to the Belavin-Polyakov (BP) skyrmion \cite{BelPolJETP75} with the
spin components 
\begin{align}
s_{x} & =2\lambda\frac{r(\cos\phi\cos\gamma-\sin\phi\sin\gamma)}{r^{2}+\lambda^{2}},\nonumber \\
s_{y} & =2\lambda\frac{r(\sin\phi\cos\gamma+\cos\phi\sin\gamma)}{r^{2}+\lambda^{2}},\nonumber \\
s_{z} & =\frac{\lambda^{2}-r^{2}}{\lambda^{2}+r^{2}}\label{skyrmion-components}
\end{align}
written as functions of polar coordinates, $r=\sqrt{x^{2}+y^{2}}$
and $\phi$, in the $xy$ plane. Here $\lambda$ is the skyrmion size
and $\gamma$ is the chirality angle. It is the energy minimum of
the first term in Eq.\ (\ref{E-continuous}) within the homotopy
class $Q=1$. That energy minimum is independent of $\lambda$ and
$\gamma$ and equals $E=4\pi JS^{2}$.

Interactions weaker than the ferromagnetic exchange that are present
in Eqs.\ (\ref{energy-discrete}) and (\ref{E-continuous}) deform
the BP skyrmion and make its energy depend on $\lambda$ and $\gamma$.
However, for the smallest skyrmions the extremal solution of the corresponding
equations of motion are still close to the BP shape \cite{quantum}.
This allows one to evaluate the energy of a small skyrmion by substituting
Eq.\ (\ref{skyrmion-components}) into Eq. (\ref{E-continuous}).
The result reads 
\begin{equation}
E=4\pi JS^{2}-\frac{2\pi JS^{2}}{3\lambda^{2}}-4\pi AS^{2}\lambda\sin\gamma+4\pi|H|S\lambda^{2}l(\lambda).\label{E}
\end{equation}
The second term in Eq. (\ref{E}) comes from the discreteness of the
lattice. It favors the shrinkage of the skyrmion \cite{cai12}. The
third and the fourth terms come from the integration of the DMI and
Zeeman interactions. The factor $l(\lambda)$ depends logarithmically
on $\delta_{H}/\lambda$ and $L/\lambda$, with a shorter of the two
lengths, $\delta_{H}=\sqrt{JS/|H|}$ or the size of the system $L$,
dominating the dependence. For $\lambda\ll\delta_{H}$ a more accurate
expression is given by \cite{quantum} 
\begin{equation}
l(\lambda)=\ln(1.5+0.68\delta_{H}/\lambda)\label{log}
\end{equation}
that will be used below. Stabilizing magnetic field is applied opposite
to the magnetic moment of the skyrmion, making the sign of the Zeeman
term positive.

\begin{figure}
\centering{} \includegraphics[width=1\columnwidth]{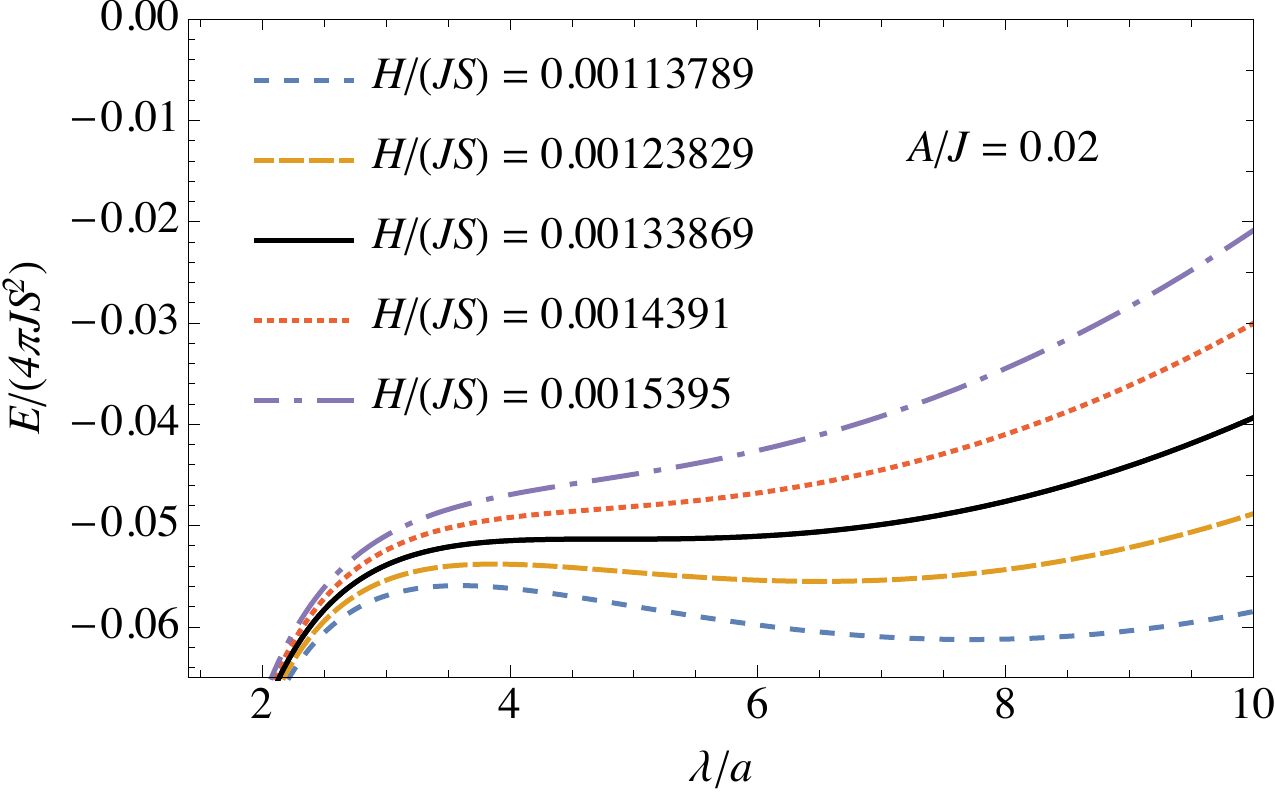} \caption{Energy vs size for an isolated metastable skyrmion at various fields.
The skyrmion size $T=0$ corresponds to the metastable energy minimum.
As the field increases, the minimum eventually disappears and the
skyrmion becomes absolutely unstable against the collapse.}
\label{Fig-e-vs-r} 
\end{figure}

The energy in Eq.\ (\ref{E}) is minimized by the Bloch-type skyrmion
with $\gamma=\pi/2$ for $A>0$. It is plotted in Fig.\ \ref{Fig-e-vs-r},
for different fields, as function of the skyrmion size $\lambda$.
On increasing the field the energy minimum shifts towards smaller
$\lambda$. At $|H|>H_{{\rm c}}$ the minimum no longer exists, making
skyrmions of size $\lambda<\lambda_{{\rm crit}}$ absolutely unstable
against the collapse.

\begin{figure}
\centering{} \includegraphics[width=1\columnwidth]{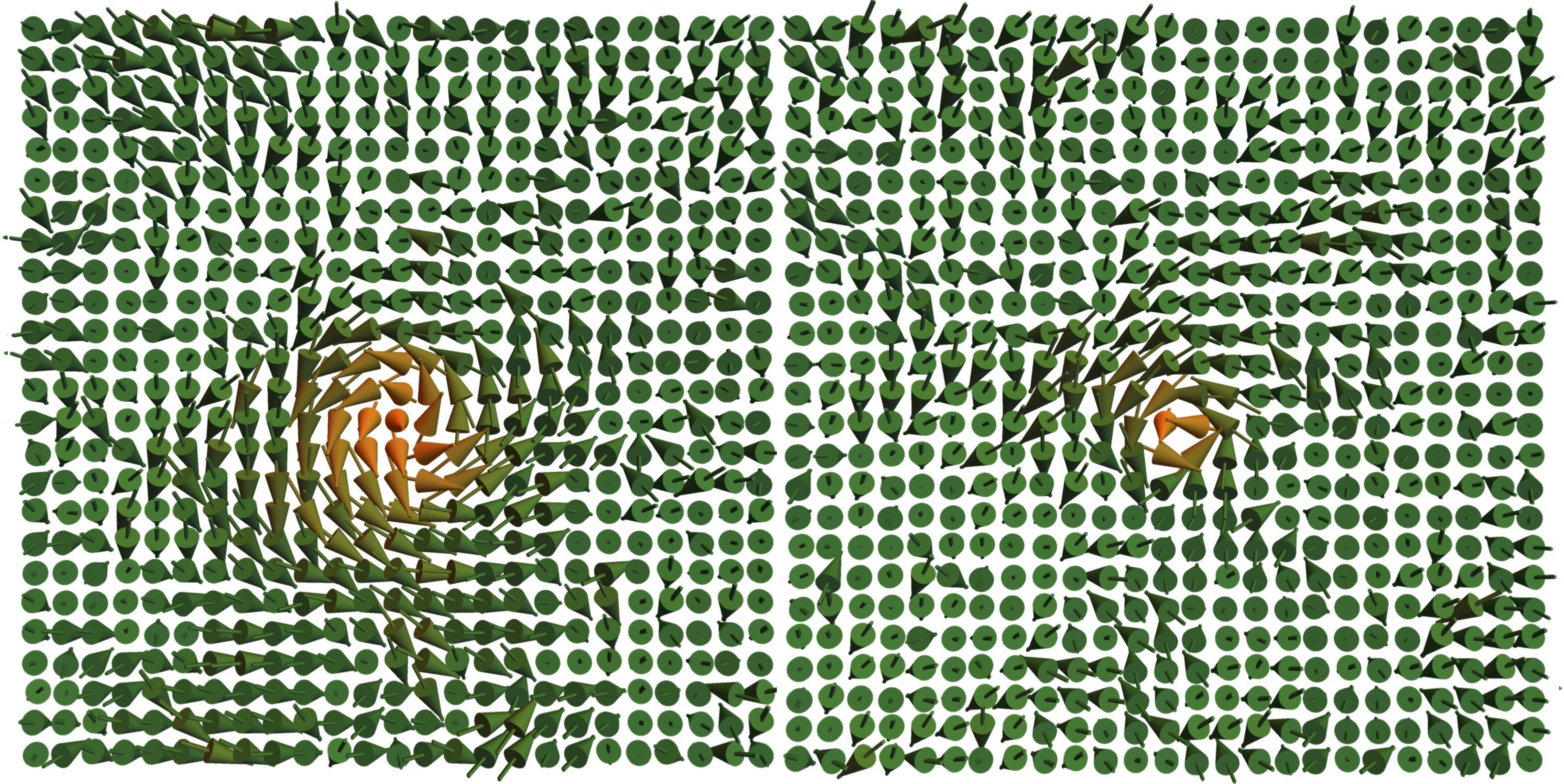} \caption{Stages of skyrmion thermal collapse. From left to right: The skyrmion
shrinks while preserving its Bloch shape with the counterclockwise
rotation. Color code: green/orange = negative/positive $s_{z}$. }
\label{Fig-collapse} 
\end{figure}

With the logarithmic accuracy $|H_{{\rm crit}}|/(JS)\sim(A/J)^{4/3}$
and $\lambda_{{\rm crit}}/a\sim(J/A)^{1/3}$. Close to the critical
field the energy barrier in Eq.\ (\ref{E}) scales as 
\begin{equation}
\frac{U}{JS^{2}}\propto\left(\frac{A}{J}\right)^{2/3}\left(1-\frac{|H|}{H_{{\rm crit}}}\right)^{n},\qquad n=\frac{3}{2}.\label{U_critical_region}
\end{equation}

\begin{figure}
\begin{centering}
\includegraphics[width=1\columnwidth]{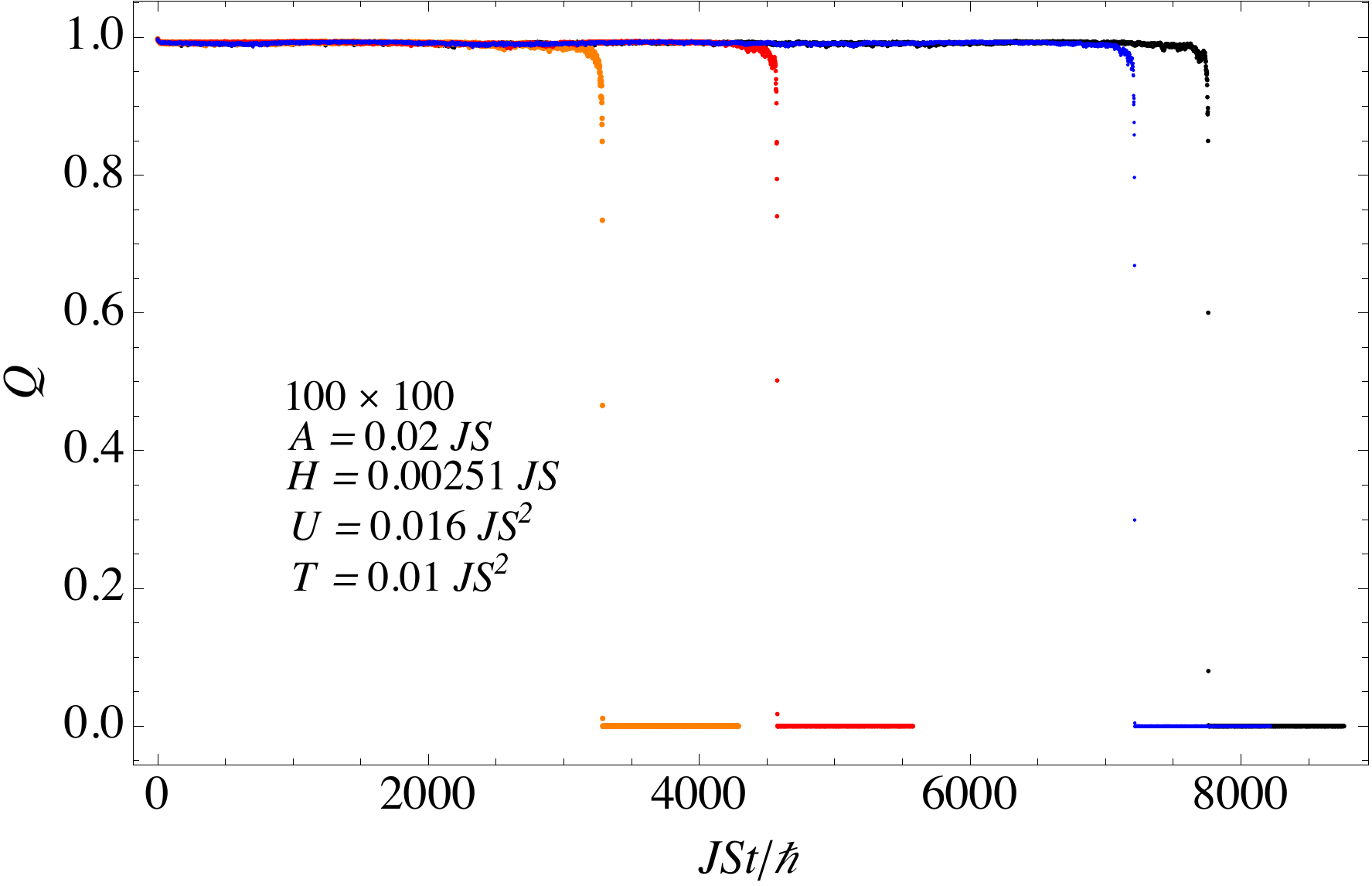}
\par\end{centering}
\caption{Different events of skyrmion collapse detected by the time dependence
of the topological charge $Q(t)$.}
\label{Fig-Qt}
\end{figure}

\section{Dynamics and numerical methods}

\label{numerical-methods}

We approach the problem of skyrmion thermal collapse by solving the
LLL equation on a 2D lattice. The time evolution of each spin on the
lattice satisfies \cite{ec-book,LandauLifshitz35,brown1963} 
\begin{equation}
\dot{\textbf{s}}_{i}=\gamma[\textbf{s}_{i}\times(\textbf{B}_{\mathrm{eff},i}+\boldsymbol{\zeta}_{i})]-\gamma\alpha[\textbf{s}_{i}\times(\textbf{s}_{i}\times\textbf{B}_{\mathrm{eff},i})],\label{LLL}
\end{equation}
where $\gamma=g\mu_{B}/\hbar$ is the gyromagnetic ratio, $\alpha$
is a dimensionless damping constant, $\textbf{B}_{\mathrm{eff},i}$
is the effective magnetic field induction acting in the $i$-th spin
and given by $\mu_{0}\textbf{B}_{\mathrm{eff},i}=-\partial\mathcal{H}/\partial\textbf{s}_{i}$,
$\mu_{0}=g\mu_{B}S$ is the magnetic moment associated with ${\bf s}_{i}$,
and $\boldsymbol{\zeta}_{i}$ is the stochastic field due to thermal
fluctuations. The equilibrium solution of Fokker-Planck equation corresponding
to this stochastic differential equation should be the Boltzmann distribution,
which requires that the stochastic field components satisfy the white-noise
condition

\begin{equation}
\langle\zeta_{\nu,i}(t),\zeta_{\beta,j}(t')\rangle=\frac{2\alpha k_{B}T}{\gamma\mu_{0}}\delta_{ij}\delta_{\nu\beta}\delta(t-t'),
\end{equation}
where $T$ is the temperature.

To speed up the numerical integration of the LLL equation, one can
replace the continuous white noise by the pulse noise with the period
$\Delta t$ \cite{garanin17,garanin18}. Noiseless evolution during
the interval $\Delta t$ between the pulses can be computed by an
efficient high-order ODE solver such as fourth-order Runge-Kutta method
with the integration step $\delta t\ll\Delta t$ for weak damping. 

Noise pulses rotate each spin by the angle

\begin{equation}
\boldsymbol{\phi}_{i}=\sqrt{\Lambda_{N}\Delta t}\textbf{G}_{i},
\end{equation}

\noindent where 
\begin{equation}
\Lambda_{N}\equiv\frac{2\gamma\alpha k_{B}T}{\mu_{0}}=\frac{2\alpha k_{B}T}{\hbar S}
\end{equation}
 is the so-called Néel attempt frequency and $\textbf{G}_{i}$ is
a three-component vector, each component being a realization of a
normal distribution with dispersion $\sigma=1$, so that $\left\langle G_{i\nu}G_{j\beta}\right\rangle =\delta_{ij}\delta_{\nu\beta}$.
With $\boldsymbol{\varphi}=\varphi\mathbf{n}$ and $|\mathbf{n}|=1$
the spin rotation formula reads
\begin{equation}
\mathbf{s}'=\mathbf{s}\cos\varphi+\left(\mathbf{n}\times\mathbf{s}\right)\sin\varphi+\mathbf{n}\left(\mathbf{s\bullet n}\right)\left(1-\cos\varphi\right).\label{eq:s_rotation}
\end{equation}
The applicability of the pulse-noise model requires that both random
rotation angles and the non-thermal relaxation angles during the interval
$\Delta t$ between the pulses be small:
\begin{equation}
\phi\sim\sqrt{\Lambda_{N}\Delta t}\ll1,\qquad\gamma\alpha B_{\mathrm{eff}}\Delta t\ll1.\label{conditions}
\end{equation}
These conditions set an upper limit on $\Delta t$. However, for $\alpha\ll1$
and low temperatures, $\Delta t$ can be much larger than the noiseless
integration step $\delta t$ that is, in turn, limited by the exchange
interaction, as the stability of typical ODE solvers requires $\delta t\lesssim0.2\hbar/(JS)$.
For $\delta t\ll\Delta t$ most of the computer time is spent on computing
the noiseless evolution, whereas random spin rotations take negligible
computer time. Testing of different choices for $\Delta t$ and $\delta t$
for a many-spin system can be found in Sec. IV of Ref. \cite{garanin17}.
Those who wish to implement the pulse-noise routine, can test it at
equilibrium comparing to Monte Carlo.

\begin{figure}
\centering{} \includegraphics[width=1\columnwidth]{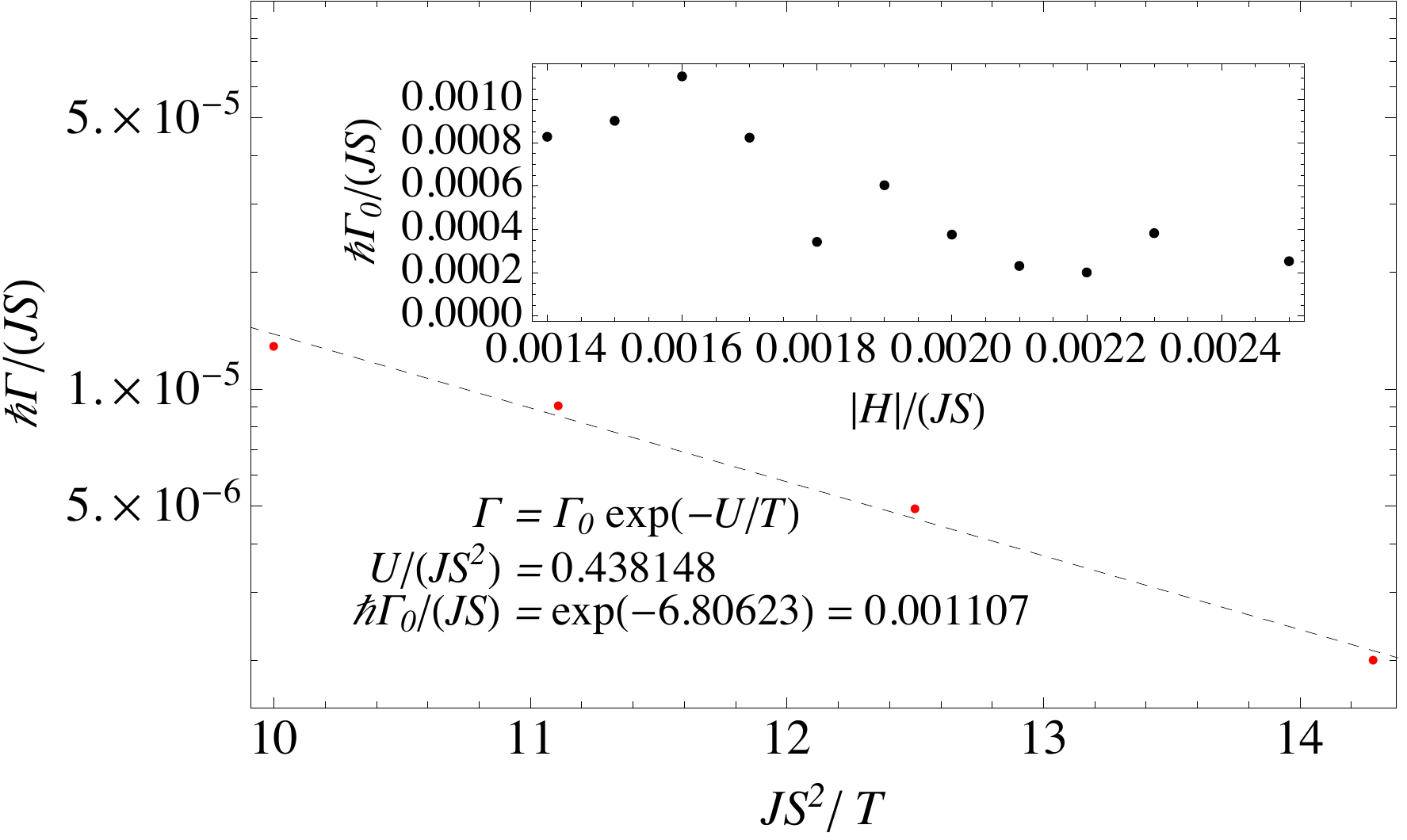}
\caption{Arrhenius dependence of the collapse rate $\Gamma$ on temperature
at $|H|/(JS)=0.0016$. Inset: Pre-exponential factor in the Arrhenius
law for the collapse rate law at various magnetic fields.}
\label{Fig-gamma-vs-T} 
\end{figure}

Our computations were done on a $100\times100$ lattice with periodic
boundary conditions. The temperature is measured in the energy units,
$k_{B}\Rightarrow1$. All energies have been measured in units of
$JS^{2}=1$, while times have been measured in units of $\hbar/(JS$).
The DMI constant $A$ was set to $0.02J$. The noise-free time interval
$\Delta t=2$ and the noise-free integration step $\delta t=0.2$
with the fourth-order Runge-Kutta method were used. Most of the computations
were performed with the damping parameter $\alpha=0.01$, so the conditions
in Eq. (\ref{conditions}) were satisfied. 

First, the metastable energy minimum for the skyrmion was found by
the energy minimization at $T=0$. This computation also yields the
numerical value of the critical field for the skyrmion collapse, $H_{{\rm crit,N}}$.
Then, the thermal noise was switched on and the LLL was solved until
the maximum time of $t_{\max}=10^{6}$. Such computations were run
in parallel 300 times for each set of parameters on a 40-node computing
cluster. The skyrmion collapse was detected by the change of the topological
charge $Q$ from the value close to 1 to the value close to zero (thermal
fluctuations wash out the value of $Q$ only slightly). We used the
four-point lattice approximation for the first derivatives in the
formula for $Q$, Eq. (\ref{Q}). After the skyrmion collapse or after
reaching the maximal time, the particular computation was terminated
and a new one began. The collected statistics of collapse times was
used to extract the collapse rate $\Gamma$ with the help of the new
algorithm that does not require that all skyrmions collapse (see the
Appendix of Ref. \cite{garanin18}). Finally, the obtained data for
$\Gamma$ were fit to the Arrhenius law, $\Gamma=\Gamma_{0}\exp(-U/T)$,
to extract the exponent $U$ and the prefactor $\Gamma_{0}$. 

%%%%%%%%%%%%%%%%%%%%%%%%%%%%%%%%%%%%%%%%%%%%%%%%%%%%%%%%%%%%% Results

\section{Numerical results and analysis}

\label{thermal-collapse}

Snapshots of spin configurations at an elevated temperature in Fig.
\ref{Fig-collapse} show a significant thermal disordering of spins
and deformation of the skyrmion seen in the middle. In spite of this,
detection of the skyrmion by the topological charge $Q$ is quite
reliable. A sample of time dependences $Q(t)$ showing the skyrmion
collapse is given in Fig. \ref{Fig-Qt}. 

The extracted dependence of the skyrmion collapse rate $\Gamma$ on
temperature together with the magnetic-field dependence of the prefactor
$\Gamma_{0}$ is shown in Fig. \ref{Fig-gamma-vs-T}. The former is
a straight line in the Arrhenius plot that confirms the expected Arrhenius
dependence $\Gamma=\Gamma_{0}\exp(-U/T)$. The prefactor $\Gamma_{0}$
in the inset decreases with increasing $|H|$. Whereas there is no
comprehensive theory yet, one can expect that the skyrmion is relaxing
by the energy exchange with spin waves. Increasing $|H|$ pushes the
gap in the spin-wave spectrum up, whereas the frequency of the skyrmion
oscillations near its energy minimum decreases as the minimum flattens.
Thus, the frequency mismatch between the skyrmion and spin waves increases
that suppresses the energy exchange between them. 

Table \ref{Tab-damping_dependence} shows the values of the barrier
and the prefactor for different values of $\alpha$ computed for $|H|/(JS)=0.0019$.
Whereas the same values of $U$ are natural and only provide a self-consistency
check of our computations, the absence of the $\alpha$-dependence
of the prefactor (within the numerical accuracy of the computations)
indicates the dominance of the intrinsic damping mechanism in our
system. 

\begin{table}
\begin{centering}
\begin{tabular}{|c|c|c|c|c|c|c|}
\hline 
$\alpha$ & 0.01 & 0.02 & 0.04 & 0.06 & 0.08 & 0.1\tabularnewline
\hline 
\hline 
$U/(JS^{2})$ & 0.235  & 0.216  & 0.223  & 0.210  & 0.175  & 0.210 \tabularnewline
\hline 
$\hbar\Gamma_{0}/(JS)\times10^{3}$ & 0.604  & 0.687  & 1.03  & 1.04 & 0.760 & 1.36\tabularnewline
\hline 
\end{tabular}
\par\end{centering}
\caption{The energy barrier $U$ and prefactor $\Gamma_{0}$ of the skyrmion
collapse are independent of the damping constant $\alpha$ within
the numerical accuracy (data for $|H|/(JS)=0.0019)$. }

\label{Tab-damping_dependence}
\end{table}

\begin{figure}
\centering{}\includegraphics[width=1\columnwidth]{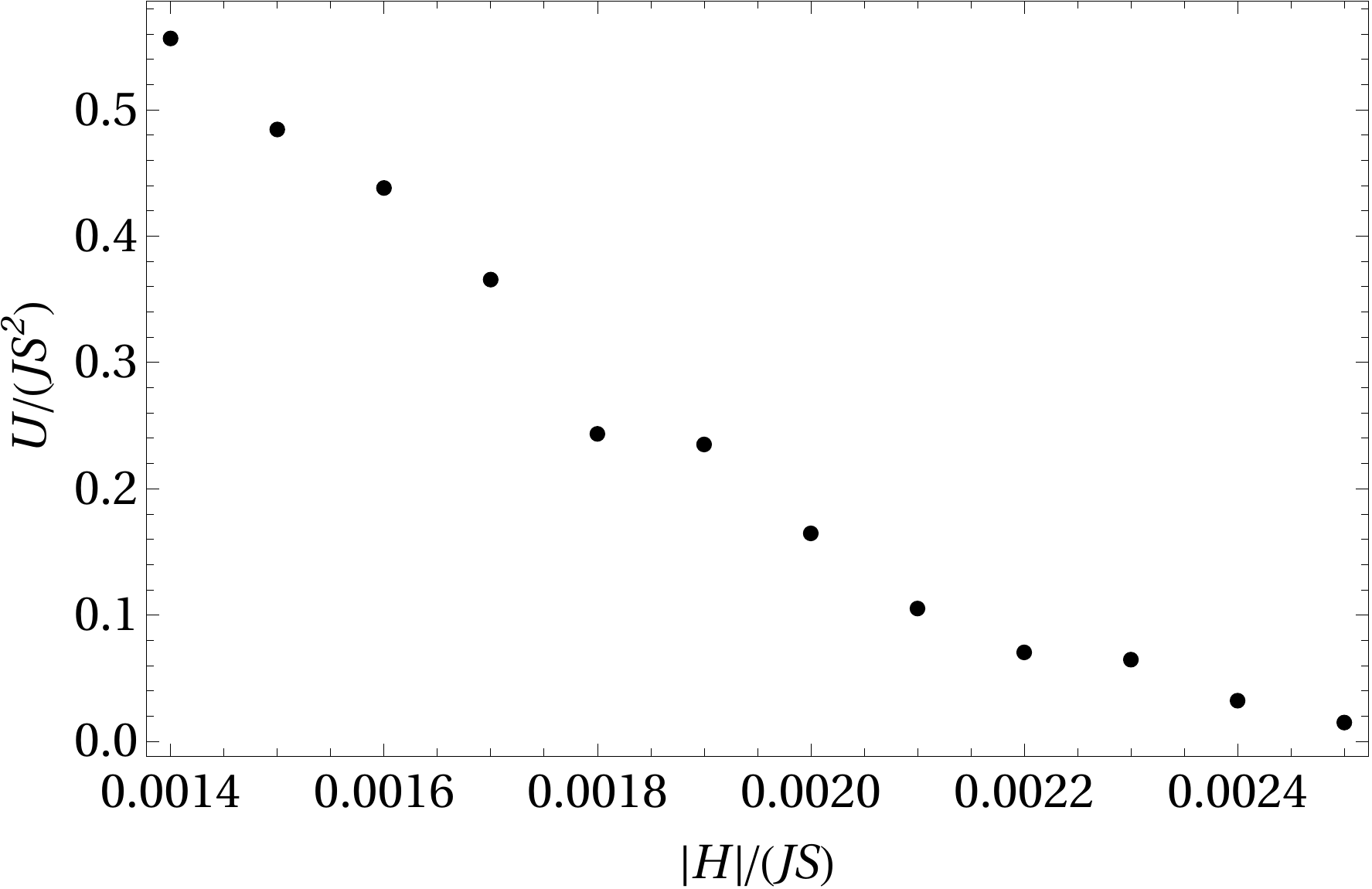} \caption{Energy barrier for the skyrmion collapse $U$ vs the external magnetic
field $|H|$.}
\label{Fig-energy-barriers} 
\end{figure}

\begin{figure}
\centering{} \includegraphics[width=1\columnwidth]{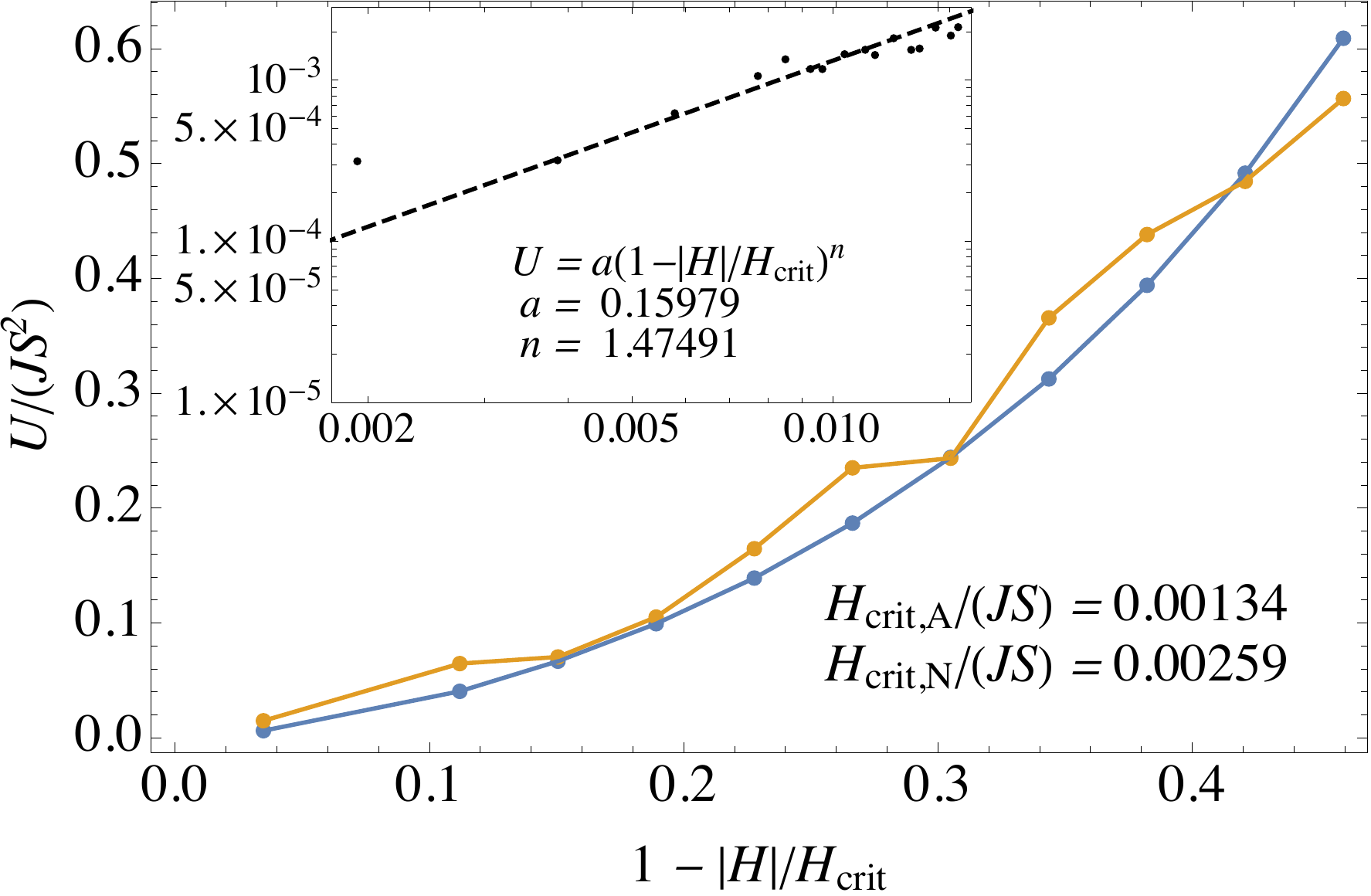}
\caption{Field dependence of the energy barrier for numerical (orange) and
analytical (blue) models compared for the same displacements from
their respective critical fields. The barrier dependence of the analytical
model has been calculated from Eq. (\ref{E}). Inset: Numerically
found dependence of the energy barrier on the deviation of the magnetic
field from the critical point (log-log scale). }
\label{Fig-barriers-plot} 
\end{figure}

The energy barriers obtained from the Arrhenius law at various magnetic
fields are presented in Fig.\ \ref{Fig-energy-barriers}. It is interesting
to compare the values determined numerically to the ones found by
computing the difference between the maximum and minimum of the energy
curve in Eq. (\ref{E}). Notice that the analytical model assumes
quasi-continuity of the spin-field. Consequently, the results obtained
analytically for the continuous spin field and numerically with the
discrete model are expected to be close to each other only when spins
rotate by a small angle from one lattice site to the neighboring one.
This is certainly not the case for small skyrmions, see Fig. \ref{Fig-collapse}.
Nevertheless, a qualitative agreement should be expected. While one
cannot directly compare values at the same external field in both
models, comparison of the barriers at the same displacement from the
critical field offers some interesting agreement. The critical field
in the discrete numerical model is $H_{{\rm crit,N}}=0.00259JS$,
while the one in the analytical model (determined by the inflection
point in the energy vs $\lambda$ in Fig.\ \ref{Fig-e-vs-r}) is
$H_{{\rm crit,A}}=0.00134JS$. Away from the critical field, we show
that the barriers obtained by the two methods are approximately the
same when computed as functions of $1-|H|/H_{{\rm crit}}$, see Fig.\ \ref{Fig-barriers-plot}.

The typical values of the pre-exponential factor in the Arrhenius
law in problems of thermal collapse range from $10^{9}$ to $10^{11}$Hz,
as found by spin dynamics computations \cite{rohart16,rozsa16}. Our
results shown in the inset of Fig. \ref{Fig-gamma-vs-T} are in accordance
with this after conversion to natural units. Using the typical value
$J=10^{3}$K and $S=1$, one obtains the values of the order from
$10^{9}$ to $10^{11}$Hz for $\Gamma_{0}$ in Fig. \ref{Fig-gamma-vs-T}.
The magnetic-field dependence of the prefactor is similar to that
observed in the experiment \cite{wild17}.

A temperature-dependent analysis of the prefactor has been given in
Ref. \cite{bessarab18} in which the authors used the presence of
Goldstone modes to derive the influence of the temperature. However,
our numerical results do not show any dependence of the prefactor
on the temperature and phenomenological damping.

Having derived in Section \ref{model} the analytical relationship
between the energy barrier and the small deviation from the critical
field, Eq. (\ref{U_critical_region}), we can compare it with our
numerical results. Fitting to Eq. (\ref{U_critical_region}) yields
$n=1.47$ that is close to the analytical value $n=3/2$. This result
is interesting because of the 3/2 power-law dependence of the energy
barrier has been observed in resonant tunneling structures\cite{oleg05,oleg03}.

%%%%%%%%%%%%%%%%%%%%%%%%%%%%%%%%%%%%%%%%%%%%%%%%%%%%%%%%%%%%%%% Discussion

\section{Discussion}

\label{discussion}

The values used in the computations were $S=1$ for the spin, $A/J=0.02$
for the ratio of DMI to exchange, and $\alpha=0.01$ for the damping
parameter. The value of the exchange constant $J=10^{3}$K was chosen
for estimates of the Arrhenius prefactor. These are typical parameters
of magnetic materials used in the experiments with skyrmions.

With the parameters chosen, the energy barriers for the skyrmion collapse
ranged from a few meV to a few tens of meV depending on the field.
Such a strong dependence of the energy barrier on the magnetic field
in the same ball park has been observed experimentally in a skyrmion
lattice \cite{wild17}.

The previous work has suggested that the skyrmion collapse rate may
be affected by changes in the pre-exponential factor alongside with
the changes in the energy barrier \cite{wild17,bessarab18}. We found
evidence of this effect on our simulations. This motivates further
research on how the pre-exponential factor impacts the skyrmion lifetime
at elevated temperatures.

Since we considered a system with periodic boundary conditions, the
only way for the skyrmion to vanish (besides quantum decay \cite{quantum}
was its spontaneous over-barrier shrinking due to thermal fluctuations
to the size below which it collapses. This is a generic problem of
skyrmion thermal collapse that must be of interest for experiments
with skyrmions at elevated temperatures, as well as for applications
of skyrmions in logic devices.

It has been shown previously \cite{bessarab18} that in systems with
boundaries there is a range of external fields for which skyrmion
escape through the boundary has a lower energy barrier than that for
the internal collapse. This possibility has sparked interest in methods
of suppressing such escape by, e.g., altering the DMI at the boundary
\cite{stosic17}. It has been shown that skyrmion stability can also
be increased by the exchange frustration \cite{vonmalottki17}. Our
method can be easily generalized for the study of these effects as
well as for the studies of bilayers \cite{koshibae17} and slabs.

\section{Conclusion}

We have numerically modeled the thermal collapse of a skyrmion stabilized
by the ferromagnetic exchange, DMI, and external magnetic field in
a two-dimensional lattice. Using the stochastic Landau-Lifshitz equation
and approximating thermal fluctuations through a pulse-noise model,
we have computed the time evolution of each spin in the lattice. In
this way, we modeled the evolution of the skyrmion in time. We used
these results to calculate the energy barrier of a skyrmion, and confirmed
them with an analytical model. Additionally, order of magnitude agreement
of energy barriers with experiment indicate that modelling the real
time dynamics of skyrmions using the pulse-noise method is an effective
way of modeling skyrmions.

\section{Acknowledgements}

This work has been supported by the grant No. DEFG02-93ER45487 funded
by the U.S. Department of Energy, Office of Science.

%%%%%%%%%%%%%%%%%%%%%%%%%%%%%%%%%%%%%%%%%%%%%%%%%%%%%%%%%%%%%%%%%%% bibliography

\end{document}